\documentclass[12pt,a4paper,twoside]{article}

\usepackage[T1]{fontenc}
\usepackage[latin1]{inputenc}
\usepackage{epsfig}
\usepackage{amssymb}
\usepackage{latexsym}
\usepackage{color}
\usepackage{times}
\usepackage{theorem}
\usepackage{amsfonts,amsmath}

\newcommand{\R}{{\mathord{\mathbb R}}}
\newcommand{\Z}{{\mathord{\mathbb Z}}}
\newcommand{\N}{{\mathord{\mathbb N}}}
\newcommand{\C}{{\mathord{\mathbb C}}}

\newcommand{\T}{{\mathord{\mathbb T}}}
\newcommand{\mA}{\mathcal A}

\newcommand{\mH}{{\mathcal H}}

\newcommand{\mL}{{\mathcal L}}
\newcommand{\mS}{{\mathcal S}}

\newcommand{\mO}{{\mathcal O}}

\newcommand{\mF}{{\mathcal F}}

\newcommand{\s}{\scriptsize}

\newcommand{\sign}{{\rm sign}}
\newcommand{\tr}{{\rm tr}}

\newcommand{\hh}{{{\mathfrak h}}}
\newcommand{\Spin}{{\mathfrak S}}
\newcommand{\CAR}{{\mathfrak F}}

\newcommand{\ch}{{\rm ch}}
\newcommand{\sh}{{\rm sh}}
\newcommand{\thh}{{\rm th}}

\newcommand{\diag}{{\rm diag}}

\newtheorem{thm}{Theorem}

\newtheorem{lemma}[thm]{Lemma}

\newtheorem{corollary}[thm]{Corollary}

\begin{document}
\pagestyle{myheadings}
\markboth{W.H. Aschbacher}{Non-zero entropy density in the XY chain out of equilibrium}

\title{Non-zero entropy density in the XY chain out of equilibrium}

\author{Walter H. Aschbacher\footnote{aschbacher@ma.tum.de}
\\ \\
Technische Universit\"at M\"unchen \\ 
Zentrum Mathematik, M5\\
85747 Garching, Germany}

\thispagestyle{empty}
\date{}
\maketitle

\begin{abstract}

The von Neumann entropy density of a block of $n$ spins is proved to be non-zero for large $n$ in the non-equilibrium steady state of the XY chain constructed by coupling a finite cutout of the chain to the two infinite parts to
its left and right which act as thermal reservoirs at different
temperatures. Moreover, the non-equilibrium density is shown to be strictly greater than the density in thermal equilibrium.

\end{abstract}

{\bf Mathematics Subject Classifications (2000).} 46L60, 47B35, 82C10, 82C23.

{\bf Key words.} Non-equilibrium steady state, XY chain, von Neumann entropy, Toeplitz operators.

\section{Introduction}

In this letter, we study the large $n$ asymptotic behavior of the von Neumann entropy 
\begin{eqnarray}
\label{entanglement}
{\mathcal S}^{(n)}=-\tr\,(\rho^{(n)}\log \rho^{(n)})
\end{eqnarray}
of the reduced density matrix $\rho^{(n)}$ which is the restriction to a subblock of $n$ neighboring spins of the non-equilibrium steady state $\omega_+$ from \cite{AP03} on the anisotropic XY chain in an external magnetic field whose formal Hamiltonian is specified by
\begin{eqnarray}
\label{XY_Hamiltonian}
 H=-\frac{1}{4}
 \sum_{x\in\Z}\left\{(1+\gamma)\,\sigma_1^{(x)}\sigma_1^{(x+1)}
 +(1-\gamma)\,\sigma_2^{(x)}\sigma_2^{(x+1)}+2\lambda
 \, \sigma_3^{(x)}\right\}.
\end{eqnarray}
Here, $\sigma_j^{(x)}$, $j=1,2,3$,  denote the Pauli matrices at site $x\in\Z$ (see \eqref{PauliMatrices} below), the parameter $\gamma\in (-1,1)$ describes the anisotropy of the spin-spin coupling, and $\lambda\in\R$ stands
for the external magnetic field. In \cite{AP03} (and, for $\gamma=\lambda=0$,  already in \cite{AH} by a different method), the non-equilibrium steady state $\omega_+$ has been constructed in a setting which has become to serve as paradigm in non-equilibrium quantum statistical mechanics: a finite cutout of the chain, the ``small'' system, is coupled to the two infinite parts to its left and right acting as thermal reservoirs at different temperatures (see Section \ref{sec:setting} for a more detailed description).\\
Since the discovery of their ideal thermal conductivity in such states, quasi-one-dimensional Heisenberg  spin-$1/2$ systems, made from different materials, have been intensively investigated experimentally (see for example \cite{SFGOVR00, SGOVR01}; the materials SrCuO$_2$ and Sr$_2$CuO$_3$ are considered to be among the best physical realizations of one-dimensional XYZ Heisenberg models) and theoretically (see for example \cite{CZP95, Z99, ZNP97}).
Not only this  highly unusual transport property motivates the theoretical study of correlations in such non-equilibrium models, but the XY chain also represents one of the simplest non-trivial testing grounds for the development of general ideas in rigorous non-equilibrium quantum statistical mechanics.

With this motivation in mind, we construct the reduced density matrix $\rho^{(n)}$ of $\omega_+$ as in \cite{LRV04, VLK03, JK04} and prove that the von Neumann entropy \eqref{entanglement} is asymptotically linear for large block size $n$. Moreover, we show that its non-equilibrium limit density  is strictly greater than the limit density in thermal equilibrium.

For further applications of the von Neumann entropy, see Remark 7.

\vspace{-1cm}

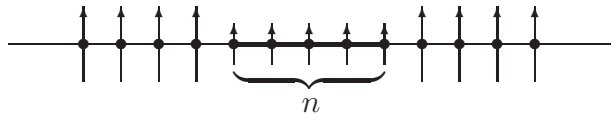
\begin{figure}[h!]
\setlength{\unitlength}{1cm}
\begin{center}
\begin{picture}(7,2)
\multiput(0,0)(0.5,0){13}{\circle*{0.15}}
\multiput(0,-0.5)(0.5,0){4}{\vector(0,1){1}}
\multiput(4.5,-0.5)(0.5,0){4}{\vector(0,1){1}}
\multiput(2,-0.27)(0.5,0){5}{\vector(0,1){0.55}}
\put(-1,0){\line(1,0){3}}
\put(4,0){\line(1,0){3}}
\put(2,-0.35){$\underbrace{\hspace{2cm}}$}
\put(2.9,-0.9){$n$}
\linethickness{0.05cm}
\put(2,0){\line(1,0){2}}
\end{picture}
\end{center}
\vspace{0.3cm}
\caption{The block of $n$ neighboring spins in \eqref{entanglement}.}
\end{figure}

\section{The non-equilibrium setting for the XY chain}
\label{sec:setting}

In this section, we give a brief informal description of our non-equilibrium setting for the XY spin model on $\Z$. We refer to \cite{AJPP06, AP03, JP02} for a precise formulation within the framework of $C^\ast$ algebraic quantum statistical mechanics (see also Remark 1 below).

Consider the XY chain described by the Hamiltonian \eqref{XY_Hamiltonian} and remove the bonds at any two sites. Doing so, the  initial spin chain divides into a compound of three noninteracting subsystems, a left infinite chain $\Z_L$, a finite piece  $\Z_\Box$, and a right infinite chain $\Z_R$.  This configuration is what we call the free system whose Hamiltonian
$$
H_0=H_L+H_\Box+H_{_R}
$$
is built on the parts $\Z_L$, $\Z_\Box$, and $\Z_R$ according to \eqref{XY_Hamiltonian}. The infinite pieces $\Z_L$ and $\Z_R$ will play the role
of thermal reservoirs to which the finite system $\Z_\Box$ is
coupled by means of the perturbation $H-H_0$. In contrast, the  configuration
described by $H$, i.e. the original XY
chain on the whole of $\Z$,  is considered to be the perturbed system. In order to construct a non-equilibrium steady state $\omega_+$ in the sense of \cite{R00}, we choose the
initial state $\omega_0$ to be composed of thermal equilibrium states on the spins on $\Z_L$, $\Z_\Box$, and $\Z_R$, with inverse temperatures $\beta_L$, $\beta_\Box$, and $\beta_R$, respectively.\\
It is well-known that the XY spin model can be described in terms of a model of free fermions with annihilation and creation operators $b_x$, $b_x^\ast$, $x\in\Z$, satisfying the canonical anticommutation relations (CAR), 
\begin{eqnarray}
\label{car}
\{b_x,b_y\}=0,\quad \{b_x^\ast,b_y\}=\delta_{xy},
\end{eqnarray}
where $\{A,B\}=AB+BA$, the operator $b_x^\ast$ is the adjoint of $b_x$, and $\delta_{xy}$ is the Kronecker symbol. This description is achieved with the help of the Jordan-Wigner transformation \cite{JW28}, 
\begin{eqnarray}
\label{def:jw}
b_x=T\pi^{(x)}(\sigma_1^{(x)}-i\sigma_2^{(x)})/2,
\end{eqnarray}
where $\pi^{(x)} = \sigma_3^{(1)} \ldots\sigma_3^{(x-1)}$ for $x>1$,
$\pi^{(1)}=1$, and $\pi^{(x)} = \sigma_3^{(x)} \ldots\sigma_3^{(0)}$
for $x<1$ (the spin $\sigma_j^{(x)}$ at site $x$, $j=1,2,3$,  is given by $...\otimes 1_2\otimes 1_2\otimes \sigma_j\otimes  1_2\otimes 1_2...$, and $1=...\otimes 1_2\otimes 1_2\otimes...$; $1_m$ denotes the identity matrix on $\C^m$). Here, $\sigma_0,...,\sigma_3$ is the Pauli basis of $\C^{2\!\times\!2}$,
\begin{eqnarray}
\label{PauliMatrices} \sigma_0=\left[\begin{array}{cc}1 & 0\\ 0&
1\end{array}\right], \quad \sigma_1=\left[\begin{array}{cc}0 & 1\\
1& 0\end{array}\right],\quad \sigma_2=\left[\begin{array}{cc}0 &
-i\\ i & 0\end{array}\right],\quad
\sigma_3=\left[\begin{array}{cc}1 & 0\\ 0& -1\end{array}\right].
\end{eqnarray}
Moreover, $T$ has the properties $T^\ast=T$, $T^2=1$, and $T\sigma_j^{(x)}T=\theta_-(\sigma_j^{(x)})$, where, for $j=1,2$,  $\theta_-$ is defined by $\theta_-(\sigma_j^{(x)})=-\sigma_j^{(x)}$ if $x\le 0$ and $\theta_-(\sigma_j^{(x)})=\sigma_j^{(x)}$ if $x\ge 1$. 

\vspace{0.5cm}

{\it Remark 1}\,\, The element $T$ stems from the $C^\ast$ crossed product extension by $\Z_2$ of the $C^\ast$ algebra $\Spin$ of the spins on $\Z$ for the two-sided chain (see  \cite{A84, AP03}; for the one-sided chain, $T$ can be dropped): The $C^\ast$ algebra $\mO$ generated by $\Spin$ and the element $T$ contains the CAR algebra $\mA$ generated by the fermions $b_x,b_x^\ast$ on $\Z$ as a $C^\ast$ subalgebra. The extension to $\mO$ of the $^\ast$-automorphisms $\theta$ on $\Spin$, defined as the rotation of the spins about the $z$-axis with angle $\pi$, $\theta(\sigma_j^{(x)})=-\sigma_j^{(x)}$ for $j=1,2$ and $\theta(\sigma_3^{(x)})=\sigma_3^{(x)}$,  yields the decomposition of $\Spin$ and $\mA$ into even and odd parts with respect to parity, $\theta(A)=\pm A$, $A\in\mO$. The even parts coincide, $\Spin_+=\mA_+$, and form a $C^\ast$ algebra (for the odd parts one has $\Spin_-=T\mA_-$ and $\Spin_-$ is merely  a Banach subspace). Since the extension to $\mO$ of the $^\ast$-automorphism groups $\tau_0$ and $\tau$ generated by $H_0$ and $H$ leave the even parts invariant, and since the unique KMS states on $\Spin$ and $\mA$ coincide on $\Spin_+=\mA_+$, the even parts only are of importance in the construction of the non-equilibrium steady state $\omega_+$ in \cite{AP03}.

\vspace{0.5cm}

Using Kato-Birman scattering theory for the construction of the wave operators on the 1-particle  Hilbert space of the Jordan-Wigner fermions \eqref{def:jw}, the non-equilibrium steady state $\omega_+$ with respect to the initial state $\omega_0$ and the time evolution $\tau^t$ generated by $H$ has been constructed in \cite{AP03} as the limit 
\begin{eqnarray*}
\omega_+(b_x)=\lim_{t\to\,+\infty}\omega_0(\tau^t(b_x))=\omega_0\circ\gamma_+(b_x),
\end{eqnarray*}
where $\gamma_+$ denotes the algebraic analogon of the wave operator on Hilbert space, the so-called M\o ller morphism on $\mA$. Moreover, it has been shown in \cite{AP03} that $\omega_+$ is a quasi-free state characterized by its 2-point operator $S$ (see Appendix \ref{app:qfs}),
\begin{eqnarray}
\label{ness}
\omega_+(B^\ast(f)B(g))=(f,Sg),
\end{eqnarray}
where, with $\hh=l^2(\Z)$,  $f,g\in \hh\oplus\hh\simeq \hh\otimes\C^2$ (and $(f,g)$ denotes the scalar product in $\hh\oplus\hh$). Here, $B$ is the linear mapping 
\begin{eqnarray}
\label{def:B}
f=(f_+,f_-)\mapsto B(f)=\sum_{x\in\Z}\,(f_+(x)\,b_x^\ast+f_-(x)\,b_x)
\end{eqnarray}
introduced in \cite{A71} in the framework of self-dual CAR algebras (see Appendix \ref{app:qfs}). It has the properties 
\begin{eqnarray}
\label{prop:B}
\{B^\ast(f),B(g)\}=(f,g),\quad B^\ast(f)=B(Jf),
\end{eqnarray} 
where $J$ is the antiunitary involution defined by $(f_+,f_-)\mapsto ({\bar f}_-,{\bar f}_+)$ (the bar denotes complex conjugation). Finally, $S$ has  been explicitly computed in \cite{AP03}. In the Fourier picture, $l^2(\Z)\otimes\C^2\simeq L^2(\T)\otimes\C^2$ \,(with $\T=\{z\in\C\,|\,|z|=1\}$),  $S$ reads 
\begin{eqnarray*} 
s(\xi)=\left(1+e^{\beta \,h(\xi)+\delta \,k(\xi)}\right)^{-1},
\end{eqnarray*}
with the temperature parameters $\beta$ and $\delta$ given by
\begin{eqnarray}
\label{def:beta_delta}
\beta=\frac{1}{2}\left(\beta_R+\beta_L\right),\quad
\delta=\frac{1}{2}\left(\beta_R-\beta_L\right).
\end{eqnarray}
The $1$-particle operators $h$ and $k$ have the form
\begin{eqnarray*}
h(\xi)=(\cos\xi-\lambda)\otimes\sigma_3-\gamma\sin\xi\otimes\sigma_2,
\quad k(\xi)=\sign(\kappa(\xi))\,\mu(\xi)\otimes\sigma_0,
\end{eqnarray*}
where the  functions $\kappa(\xi)$ and $\mu(\xi)$ are defined by
\begin{eqnarray*}
\kappa(\xi)=2\lambda\sin\xi-(1-\gamma^2)\sin
2\xi,\quad
\mu(\xi)=\left((\cos\xi-\lambda)^2+\gamma^2\sin^2\xi\right)^{1/2}.
\end{eqnarray*}
Expanding $s(\xi)$ with respect to $\sigma_0,...,\sigma_3$ from \eqref{PauliMatrices}, we find the $0$-th component of $s(\xi)=s_0(\xi)\otimes \sigma_0+\sum_{k=1}^3s_k(\xi)\otimes \sigma_k$ to look like
\begin{eqnarray}
\label{s0} s_0(\xi)&=&\frac{1}{2}-\frac{1}{2}
\,\varphi_{\delta,\beta}(\xi) \,\sign\kappa(\xi),
\end{eqnarray}
whereas $s_k(\xi)$, $k=1,2,3$, has the form
\begin{eqnarray}
\label{sk} 
s_k(\xi)=-\frac{1}{2}\,\varphi_{\beta,\delta}(\xi)\,r_k(\xi),\quad
r(\xi)=\frac{1}{\mu(\xi)}\,(0,-\gamma \sin\xi,\cos\xi-\lambda).
\end{eqnarray}
Here, we used the definition
\begin{eqnarray}
\label{def:varphi}
\varphi_{\alpha,\alpha'}(\xi)&=&\frac{\sh(\alpha \mu(\xi))}{\ch(\alpha
 \mu(\xi))+\ch(\alpha' \mu(\xi))},\quad \alpha,\alpha'\in\R.
\end{eqnarray}

In order to study the von Neumann entropy \eqref{entanglement}, we restrict $\omega_+$ to spins on the finite configuration subset $\Lambda_n=\{1,...,n\}$, $n\in\N$,  of the entire spin chain on $\Z$. Due to the Jordan-Wigner transformation \eqref{def:jw}, this is equivalent to restricting $\omega_+$ to the fermions $b_i, b_i^\ast$ at the same sites $\Lambda_n$ (on the even part of the algebra, see Remark 1). Let $\hh_n=l^2(\Lambda_n)\simeq \C^n$ be the state  space over $\Lambda_n$ \,(where here and in the following, $\hh_n$ is considered as a subspace of $\hh$ by the trivial injection). Since the Fock space $\CAR(\hh_n)$ over the $n$-dimensional state space $\hh_n$ is $2^n$-dimensional,
the CAR algebra $\mA_n\equiv\mA(\hh_n)$ over $\hh_n$ is $2^{2n}$-dimensional, $\mA_n\simeq\mL(\mF(\hh_n))\simeq\C^{2^n\!\times\! 2^n}$ (see for example \cite{AJPP06} or \cite[p.15]{BR2}; we denote by $\mL(\mH)$ the bounded operators on the Hilbert space $\mH$). Therefore, the restriction of $\omega_+$ to the spins on $\Lambda_n$ is described by a density matrix $\rho^{(n)}\in \mL(\mF(\hh_n))$ (see for example \cite[p.267]{BR2}),
\begin{eqnarray}
\label{def:rho}
\omega_+(A)=\tr(\rho^{(n)}A),\quad A\in \mA_n.
\end{eqnarray}

\section{The asymptotics of the von Neumann entropy}

In Theorem \ref{thm}, our  main result, we prove that the large $n$ asymptotics of the von Neumann entropy ${\mathcal S}^{(n)}$ from \eqref{entanglement} in the non-equilibrium steady state $\omega_+$ is, to leading order in $n$, proportional to the volume of the spin block, and we derive an explicit expression for the non-vanishing proportionality constant.  To do so, we express ${\mathcal S}^{(n)}$ with the help of the eigenvalues of a Toeplitz matrix which we will construct next. Let us first define the Majorana correlation matrix $\Omega_n\in\C^{2n\times 2n}$ by
\begin{eqnarray}
\label{def:Omega_n}
(\Omega_n)_{kl}=\omega_+(B(w^{(k)})B(w^{(l)})),\quad k,l=1,...,2n,
\end{eqnarray}
where $B(f)$ is given in \eqref{def:B}, and $w^{(k)}=(w_+^{(k)},w_-^{(k)})\in \hh_{n}\oplus\hh_n$ is specified, for $j=1,...,n$, by
\begin{eqnarray*}
w_+^{(k)}(j)&=&\left(\oplus_{1}^n\tau\right)_{k,2j}=\delta_{k,2j-1}-i\delta_{k,2j},\nonumber\\
w_-^{(k)}(j)&=&\left(\oplus_{1}^n\tau\right)_{k,2j-1}=\delta_{k,2j-1}+i\delta_{k,2j}.
\end{eqnarray*}
Moreover,  the matrix $\tau\in\C^{2\times 2}$ reads 
\begin{eqnarray}
\label{def:tau}
\tau=\left[\begin{array}{cc}1 & 1\\i & -i\end{array}\right].
\end{eqnarray}

\vspace{0.5cm}

{\it Remark 2}\,\,By virtue of \eqref{car},  the $2n$ operators $d_k=B(w^{(k)})$ are Majorana operators, i.e. 
\begin{eqnarray}
\label{def:majorana}
d_k^\ast=d_k,\quad \{d_k,d_l\}=2\delta_{kl}. 
\end{eqnarray} 

\vspace{0.5cm}

In the following, we define a suitable block Toeplitz matrix $T_n[a]$. The reader not familiar with Toeplitz theory may consult Appendix
\ref{app:Toeplitz} where the notation and the relevant facts are given. Moreover, in the proofs below, we will refer to Appendix \ref{app:qfs} for an elementary construction of a suitable fermionic basis.

\begin{lemma} 
\label{lemma:Toeplitz}
The imaginary part of the Majorana correlation matrix $\Omega_n$ from \eqref{def:Omega_n} is a skew-symmetric $2\!\times\!2$ block Toeplitz matrix $T_n[a]$ with symbol $a\in L^\infty_{2\times 2}$,
\begin{eqnarray}
\label{def:Tn}
\Omega_n=1_{2n}+iT_n[a],\quad a(\xi)=\frac{2}{i}\left[\begin{array}{cc}s_0(\xi)-\frac{1}{2} & s_2(\xi)-is_3(\xi)\\s_2(\xi)+is_3(\xi) & s_0(\xi)-\frac{1}{2} \end{array}\right],
\end{eqnarray}
where the functions $s_0,...,s_3$ are defined in \eqref{s0} and \eqref{sk}.
\end{lemma}

\vspace{0.2cm}

{\bf Proof}\,\,With the Fourier transform $F:l^2(\Z)\to L^2(\T)$ in the form  $F\!f(\xi)=\sum_{x\in\Z}f(x)\,e^{ix\xi}$, we have $Fw^{(2k-1)}_\pm=e_k$, $Fw^{(2k)}_\pm=\mp ie_k$, and $S=F^\ast\!\otimes\! 1_2\, s\,F\!\otimes\! 1_2$,  where we define $e_x(\xi)=e^{ix\xi}$. Hence, with $\eta_\pm=(1,\pm 1)\in\C^2$, we compute
\begin{eqnarray}
\label{Omega_oo}
(\Omega_n)_{2j-1,2k-1}&=&\omega_+(B(w^{(2j-1)})B(w^{(2k-1)}))\nonumber\\
&=&(w^{(2j-1)},S\, w^{(2k-1)})\nonumber\\
&=&\sum_{\alpha=0}^3(e_j\otimes \eta_+, s_\alpha e_k\otimes\sigma_\alpha\eta_+)\nonumber\\
&=&2\,(e_j,(s_0+s_1)\, e_k)\nonumber\\
&=&\delta_{jk}-\int_0^{2\pi}\!\frac{d\xi}{2\pi}\, \varphi_{\delta,\beta}(\xi)\, \sign\kappa(\xi)\,e^{-i(k-j)\xi}.
\end{eqnarray}
In the last equality only, we  used the explicit expressions of $s_0$  and $s_1$ from \eqref{s0} and \eqref{sk} (the function $\varphi_{\delta,\beta}$ is defined in \eqref{def:varphi}). In the same way, we obtain
\begin{eqnarray}
\label{Omega_oe}
(\Omega_n)_{2j-1,2k}
&=&2\,(e_j,(s_2-is_3)\,e_k)\nonumber\\
&=&-\int_0^{2\pi}\!\frac{d\xi}{2\pi}\, \frac{q(\xi)}{\mu(\xi)}\varphi_{\beta,\delta}(\xi)\, e^{-i(k-j)\xi},\nonumber\\
(\Omega_n)_{2j,2k-1}&=&2\,(e_j,(s_2+is_3)\,e_k),\nonumber\\
(\Omega_n)_{2j,2k}&=&2\,(e_j,(s_0-s_1)\,e_k),
\end{eqnarray}
where we define 
\begin{eqnarray*}
q(\xi)=-\gamma\sin\xi-i(\cos\xi-\lambda).
\end{eqnarray*} 
Hence, the matrix $T_n[a]=(\Omega_n-1_{2n})/i$ is a block Toeplitz matrix with $2\!\times\!2$ blocks, i.e. $(\Omega_n)_{k+2i,l+2i}=(\Omega_n)_{kl}$. Its symbol $a\in L^\infty_{2\times 2}$ can be read off from \eqref{Omega_oo} and \eqref{Omega_oe},
\begin{eqnarray*}
a(\xi)=i\left[\begin{array}{cc} \varphi_{\delta,\beta}(\xi)\,\sign\kappa(\xi)& \frac{q(\xi)}{\mu(\xi)}\,\varphi_{\beta,\delta}(\xi)\\ 
\frac{\bar{q}(\xi)}{\mu(\xi)}\,\varphi_{\beta,\delta}(\xi) & \varphi_{\delta,\beta}(\xi)\,\sign\kappa(\xi)\end{array}\right].
\end{eqnarray*}
Moreover, using the properties \eqref{def:majorana} of the Majorana operators for $i(T_n[a])_{kl}=\omega_+(d_kd_l)-\delta_{kl}$, we see that $T_n[a]$ is real and skew-symmetric, i.e. ($A^t$ denotes the transpose of the matrix $A$)
\begin{eqnarray*}
T_n[a]\in \R^{2n\times 2n},\quad T_n[a]^t=-\,T_n[a].
\end{eqnarray*}
This is the claim of Lemma \ref{lemma:Toeplitz}. \hfill$\Box$\\

\vspace{0.5cm}

In the next lemma, we choose a special set of fermions in $\mA_n$ stemming from the block diagonalization of $T_n[a]$ by means of real orthogonal transformation $V\in O(2n)=\{V\in\R^{2n\times 2n}\,|\,VV^t=1_{2n}\}$.

\begin{lemma}
\label{lemma:form_rho} 
There exists a set of fermions $c_i\equiv c_i^{(n)}\in\mA_n$, and numbers $\lambda_i^{(n)}\in\R$, $i=1,...,n$, such that the reduced density matrix $\rho^{(n)}$ from \eqref{def:rho} has the form
\begin{eqnarray}
\label{rho}
\rho^{(n)}=\prod_{i=1}^n\left(\frac{1+\lambda_i^{(n)}}{2}\,c_i^\ast c_i+\frac{1-\lambda_i^{(n)}}{2}\,c_ic_i^\ast\right).
\end{eqnarray}
\end{lemma}

{\bf Proof}\,\, Let $c_i$, $i=1,...,n$, be arbitrary fermions in $\mA_n$ and define operators $e_{\alpha\beta}^{(i)}$, $i=1,...,n$, $\alpha,\beta=1,2$, by
\begin{eqnarray}
\label{eij_c}
e_{11}^{(i)}=c_i^\ast c_i,\quad e_{12}^{(i)}=c_i^\ast,\quad e_{21}^{(i)}=c_i,\quad e_{22}^{(i)}=c_i c_i^\ast. 
\end{eqnarray} 
Since Lemma \ref{lemma:onb} in Appendix \ref{app:qfs} states that the $2^{2n}$ operators $\prod_{i=1}^ne_{\alpha_i\beta_i}^{(i)}$, $\alpha_1,\beta_1,...,\alpha_n,\beta_n=1,2$, constitute an orthonormal basis in $\mL(\CAR(\hh_n))$, we can write $\rho^{(n)}\in\mL(\CAR(\hh_n))$ as
\begin{eqnarray}
\label{rho1}
\rho^{(n)}=\sum_{\alpha_1,\beta_1,...,\alpha_n,\beta_n=1,2}\omega_+\left(\left(\prod_{i=1}^ne^{(i)}_{\alpha_i\beta_i}\right)^\ast\right)\prod_{j=1}^ne^{(j)}_{\alpha_j\beta_j}.
\end{eqnarray}
The goal of the following is to make a special choice for the fermions  $c_i$ such that the density matrix $\rho^{(n)}$ from \eqref{rho1} can be written in the form \eqref{rho}. For this purpose, let $V\in O(2n)$ be any real orthogonal $2n\!\times\! 2n$ matrix and choose 
\begin{eqnarray}
\label{def:bn}
c_i=B(v^{(i)}),\quad i=1,...,n,
\end{eqnarray}
where  $v^{(i)}=(v_+^{(i)},v_-^{(i)})\in \hh_n\oplus\hh_n$ is specified, for $j=1,...,n$, by 
\begin{eqnarray}
\label{def:v}
v_+^{(i)}(j)&=&\left(\left(\oplus_{1}^n\tau^{-1}\right)V\left(\oplus_{1}^n\tau\right)\right)_{2i-1,2j},\nonumber\\
v_-^{(i)}(j)&=&\left(\left(\oplus_{1}^n\tau^{-1}\right)V\left(\oplus_{1}^n\tau\right)\right)_{2i-1,2j-1},
\end{eqnarray}
and  $\tau$ is defined in \eqref{def:tau} \,(note that for the choice $V\!=\!1_{2n}$ we recover the Jordan-Wigner fermions $b_i$ from \eqref{def:jw}). Using 
\begin{eqnarray}
\label{bick}
c_i=\sum_{k=1}^{2n}\left(\left(\oplus_{1}^n\tau^{-1}\right)V\right)_{2i-1,k}d_k,
\end{eqnarray}
the Majorana relations \eqref{def:majorana}, the fact that $V\in O(2n)$, and the properties of $\tau$, we see that the operators $c_i$ are fermionic,
\begin{eqnarray*}
\{c_i,c_j\}&=&2\,((\oplus_{1}^n\tau^{-1})VV^\ast (\oplus_{1}^n{\tau^{-1}}^\ast))_{2j-1,2i-1}=(\oplus_1^n\sigma_1)_{2j-1,2i-1}=0,\\
\{c_i^\ast,c_j\}&=&2\,((\oplus_{1}^n\tau^{-1})VV^t(\oplus_{1}^n{\tau^{-1}}^t))_{2j-1,2i-1}\hspace{0.08cm}=(\oplus_1^n\sigma_0)_{2j-1,2i-1}=\delta_{ij}
\end{eqnarray*}
(note that, in general, the CAR are not satisfied if we choose $V$ to be complex unitary or complex orthogonal). Next, let us make a special choice for $V$. Since we know from Lemma \ref{lemma:Toeplitz} that $T_n[a]$  is real and skew-symmetric, there exists a matrix $Q\in O(2n)$ which transforms  $T_n[a]$ into its real canonical form (see for example  \cite[p.107]{HJ85}),
\begin{eqnarray}
\label{def:Q}
Q\,T_n[a]\,Q^t=\oplus_{j=1}^n(\lambda_j^{(n)}\,i\sigma_2)=\diag\,(\lambda_1^{(n)},...,\lambda_n^{(n)})\otimes i\sigma_2,
\end{eqnarray}
where $\pm i\lambda_j^{(n)}$, $\lambda_j^{(n)}\in\R$, are the eigenvalues of $T_n[a]$. Hence, setting $V=Q$ in \eqref{def:v}, and using \eqref{bick} and \eqref{def:Q}, we find, similarly as for the CAR above, 
\begin{eqnarray}
\label{o1}
\omega_+(c_ic_j)&=&\frac{1}{2}\left(\oplus_{k=1}^n\left(\sigma_1-\lambda_k\, i\sigma_2\right)\right)_{2i-1,2j-1}=0,\\
\label{o2}
\omega_+(c_i^\ast c_j)&=&\frac{1}{2}\left(\oplus_{k=1}^n\left(\sigma_0+\lambda_k\, \sigma_3\right)\right)_{2i-1,2j-1}=\delta_{ij}\,\frac{1+\lambda_i^{(n)}}{2}.
\end{eqnarray}
Now, let us write $\rho^{(n)}$ from \eqref{rho1} with the help of this special choice of  fermions. Since \eqref{o1} and \eqref{o2} hold, we have, from Lemma \ref{lemma:factorizing} in Appendix \ref{app:qfs}, that
\begin{eqnarray}
\label{factorizing3}
\omega_+\!\left(\left(\prod_{i=1}^ne^{(i)}_{\alpha_i\beta_i}\right)^\ast\right)=\prod_{i=1}^n\delta_{\alpha_i\beta_i}\,\omega_+(e^{(i)}_{\alpha_i\alpha_i}).
\end{eqnarray}
Therefore, plugging \eqref{factorizing3} into \eqref{rho1} and using \eqref{eij_c}, \eqref{o1}, and \eqref{o2}, we find that the reduced density matrix $\rho^{(n)}$ takes the form
\begin{eqnarray*}
\rho^{(n)}=\prod_{i=1}^n\left(\omega_+(c_i^\ast c_i)\,c_i^\ast c_i+\omega_+(c_i c_i^\ast)\,c_i c_i^\ast\right)=\prod_{i=1}^n\left(\frac{1+\lambda_i^{(n)}}{2}\,c_i^\ast c_i+\frac{1-\lambda_i^{(n)}}{2}\,c_ic_i^\ast\right),
\end{eqnarray*}
which is \eqref{rho}. \hfill$\Box$

\vspace{0.5cm}

Next, we formulate our main assertion about the asymptotic behavior of the von Neumann entropy ${\mathcal S}^{(n)}=-\,\tr\,(\rho^{(n)}\log \rho^{(n)})$ in the non-equilibrium steady state $\omega_+$ characterized by \eqref{ness} with its reduced density matrix  $\rho^{(n)}$  from \eqref{def:rho}. The temperatures $\beta$ and $\delta$ are defined in \eqref{def:beta_delta}.

\vspace{0.5cm}

\begin{thm}
\label{thm}
Let $\omega_+$ be the non-equilibrium steady state of the XY chain with reduced density matrix $\rho^{(n)}$. Then,  for the temperatures $0\le\delta<\beta<\infty$, the anisotropy $\gamma\in (-1,1)$, and the magnetic field $\lambda\in\R$, the von Neumann entropy ${\mathcal S}^{(n)}$ is asymptotically linear for $n\to\infty$, 
\begin{eqnarray*}
{\mathcal S}^{(n)}= C\,n+o(n)\quad \mbox{with}\quad C\equiv C_{\beta,\delta,\gamma,\lambda}>0,
\end{eqnarray*}
and $C$ is given in \eqref{def:C}.
\end{thm}

\vspace{0.2cm}

{\it Remark 3}\,\, The existence of the limit $\lim_{n\to\infty}\mS^{(n)}/n$ for translation invariant states of spin systems follows from the strong subadditivity property of the mapping $[1,...,n]\to\mS^{(n)}$, see for example \cite[p.287]{BR2}. In \eqref{def:C}, we derive an explicit expression for this limit and show that it is strictly positive.

\vspace{0.3cm}

{\bf Proof}\,\, The properties \eqref{ToeplitzNorm} and \eqref{def:phi_infinity} from  Appendix \ref{app:Toeplitz} imply that the spectral radius  of the Toeplitz matrix $T_n[a]$ is bounded by 
\begin{eqnarray}
\label{spec_radius}
\|T_n[a]\|\le \mbox{ess}\sup_{\hspace{-0.5cm}\xi\in \,[0,2\pi]}\|a(\xi)\|_{\mL(\C^2)}=\mbox{ess}\sup_{\hspace{-0.5cm}\xi\in \,[0,2\pi]}\,(\varphi_{\beta,\delta}(\xi)+\varphi_{\delta,\beta}(\xi))=\varrho,
\end{eqnarray}
where, due to the finiteness of the temperatures,  $\varrho$ is strictly smaller than $1$,
\begin{eqnarray}
\label{def:varrho}
\varrho=\thh(\beta_R(1+|\lambda|)/2)<1.
\end{eqnarray}
The form \eqref{rho} immediately yields the spectral representation of $\rho^{(n)}$ with the $2^n$ eigenvalues
\begin{eqnarray}
\label{ev}
\lambda_{\epsilon_1,...,\epsilon_n}=\prod_{i=1}^n\frac{1+(-1)^{\epsilon_i}\lambda_i^{(n)}}{2},\quad \epsilon_1,...,\epsilon_n\in\{0,1\}.
\end{eqnarray}
Due to \eqref{spec_radius} and \eqref{def:varrho}, we have $0<\lambda_{\epsilon_1,...,\epsilon_n}<1$. Using \eqref{ev}, $\tr\,\rho^{(n)}=1$, and the entropy function $h:(-1,1)\to\R^+$ \,(for which $h(x)=\eta((1+x)/2)$ with $\eta(x)=-x\log x-(1-x)\log(1-x)$ the Shannon entropy; see Figure 2 for $h(x)$),
\begin{eqnarray}
\label{def:h}
h(x)=-\frac{1+x}{2}\log\left(\frac{1+x}{2}\right)-\frac{1-x}{2}\log\left(\frac{1-x}{2}\right),
\end{eqnarray}
the von Neumann entropy ${\mathcal S}^{(n)}=-\tr(\rho^{(n)}\log\rho^{(n)})$ can be written in the form
\begin{eqnarray*}
{\mathcal S}^{(n)}
=-\hspace{-0.4cm}\sum_{\epsilon_1,...,\epsilon_n\in\,\{0,1\}}\hspace{-0.3cm}\lambda_{\epsilon_1,...,\epsilon_n}\,\log \lambda_{\epsilon_1,...,\epsilon_n}
=\sum_{i=1}^nh(\lambda_i^{(n)}).
\end{eqnarray*}
To determine the asymptotic behavior of ${\mathcal S}^{(n)}$, we want to make use of Szeg\"o's first limit theorem for the case of block Toeplitz matrices (see \cite[p.202]{BS99} and \eqref{szego} in Appendix \ref{app:Toeplitz}). To do so, we note that the symbol $ia\in L_{2\times 2}^\infty$ is self-adjoint (with $a$ from \eqref{def:Tn}) and that $\pm\lambda_j\in (-\varrho,\varrho)$ are the eigenvalues of the self-adjoint Toeplitz matrix $iT_n[a]=T_n[ia]$ (see \eqref{def:Q}). Moreover, with the definition ${\tilde h}(x)=h(x)$ for $x\in (-1,1)$ and ${\tilde h}(x)=0$ for $x\in \R\setminus (-1,1)$, we have ${\tilde h}\in C_0(\R)$ (the continuous functions with compact support) and hence, due to $h(-x)=h(x)$, Szeg\"o's first limit theorem \eqref{szego} implies 
\begin{eqnarray}
\label{def:C}
C_{\beta,\delta,\gamma,\lambda}&=&\lim_{n\to\infty}\frac{{\mathcal S}^{(n)}}{n}\nonumber\\
&=&\frac{1}{2}\int_0^{2\pi}\!\frac{d\xi}{2\pi}\,\tr\,{\tilde h}(ia(\xi))\nonumber\\
&=&\frac{1}{2}\int_0^{2\pi}\!\frac{d\xi}{2\pi}\,(h(\varphi_{\beta,\delta}(\xi)+\varphi_{\delta,\beta}(\xi))+h(\varphi_{\beta,\delta}(\xi)-\varphi_{\delta,\beta}(\xi)))\\
&\ge& h(\varrho)>0.\nonumber
\end{eqnarray}
This proves our theorem. \hfill $\Box$

\begin{figure}
\setlength{\unitlength}{1cm}
\begin{center}
\epsfxsize=5cm
\epsffile{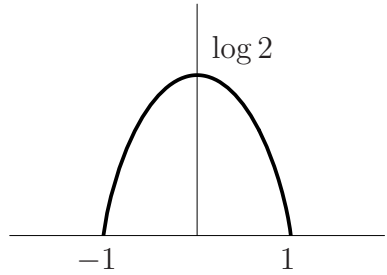}

\begin{picture}(7,2)
\put(4.6,1.6){$1$}
\put(1.9,1.6){$-1$}
\put(3.7,4.4){$\log 2$}
\end{picture}

\vspace{-1.8cm}

\caption{The entropy function $h(x)$ from \eqref{def:h}.}
\end{center}
\end{figure}

\vspace{0.5cm}

{\it Remark 4}\,\, Contrary to what one would expect naively, the singular nature of the symbol does not affect the leading order of the asymptotics of the entropy density: As soon as there is a strictly positive temperature in the system, the spectral radius of the Toeplitz operator contracts to a value strictly less than $1$. The same phenomenon has been observed for the transversal spin-spin correlations in this model, see \cite{AB06}.

\vspace{0.2cm}

{\it Remark 5}\,\, In the case $\delta=0$, the symbol of the Toeplitz operator becomes smooth. Then, second order trace formulas imply that the subleading term has the form $o(n)=const+o(1)$, see \cite[p.133, 207]{BS99}.

\vspace{0.2cm}

{\it Remark 6}\,\, Note that for $\beta\to\infty$, the right hand side of \eqref{def:C} vanishes in agreement with the findings in the literature. In this case, the logarithmic behavior of the entropy density has been derived using the asymptotics of Toeplitz operators with Fisher-Hartwig symbols, see \cite{JK04}.

\vspace{0.2cm}

{\it Remark 7}\,\, The von Neumann entropy $\mS^{(n)}$ is  widely being used as a measure of entanglement in the ground state of a variety of quantum mecanical systems. In \cite{LRV04, VLK03, JK04} for example, it has been shown that the entanglement entropy tends to a finite saturation value, whereas at a quantum phase transition (see \cite{S99}), it grows logarithmically with the size $n$ of the subsystem. These results can be related to conformal field theory computations of the geometric entropy, see, for example, \cite{S93, CW94, FPST94, HLW94}.

\vspace{0.5cm}

\begin{corollary}
For large $n$, the non-equilibrium entropy density is strictly greater than the entropy density in thermal equilibrium, i.e. $C_{\beta,\delta,\gamma,\lambda}>C_{\beta,0,\gamma,\lambda}$.
\end{corollary}

{\bf Proof}\,\, From \eqref{def:C} we have 
\begin{eqnarray*}
C_{\beta,\delta,\gamma,\lambda}=\frac{1}{2}\int_0^{2\pi}\!\frac{d\xi}{2\pi}\,(h(\thh(\beta_R\mu(\xi)/2))+h(\thh(\beta_R\mu(\xi)/2-\delta\mu(\xi))))>C_{\beta,0,\gamma,\lambda}.
\end{eqnarray*}
\hfill $\Box$

\vspace{0.5cm}

{\it Remark 8}\,\, In \cite{EZ}, following \cite{ARRS}, a non-equilibrium steady state has been constructed by imposing a current on the system (these states differ in general from the non-equilibrium steady states in the sense of \cite{R00}, see also \cite{AJPP06}). The content of the corollary is in agreement with \cite{EZ} in the sense that the von Neumann entropy increases in the current carrying state (by doubling the prefactor of the asymptotically logarithmic behavior).

\appendix

\section{Quasi-free states on self-dual CAR algebras}
\label{app:qfs}

Quasi-free states on self-dual CAR algebras have been introduced in \cite{A71}. We briefly review these notions. A self-dual CAR algebra over a Hilbert space ${\mathcal K}$ with scalar product $(\cdot,\cdot)$ and antiunitary involution $J$ is a $C^\ast$ completed $^\ast$-algebra generated by $B(f)$, $f\in {\mathcal K}$, $B^\ast(f)$, and an identity $1$ which satisfy 
\begin{eqnarray}
\label{B:car}
\{B^\ast(f),B(g)\}=(f,g),\quad B^\ast(f)=B(Jf),
\end{eqnarray}
and $B(f)$ is linear in $f$. A quasi-free state on a self-dual CAR algebra is defined  by
\begin{eqnarray}
\label{qf1}
\omega(B(f^{(1)})...B(f^{(2m-1)}))&=&0,\\
\label{qf2}
\omega(B(f^{(1)})...B(f^{(2m)}))&=&\sum_{\pi}\sign\pi\prod_{i=1}^m\omega(B(f^{(\pi(2i-1))})B(f^{(\pi(2i))})),
\end{eqnarray}
for $m\in\N$, and the sum runs over all permutations $\pi$ in the permutation group $S_{2m}$ with signature $\sign\pi$ which satisfy $\pi(2i-1)<\pi(2i),\pi(2i+1)$ (the pairings of $\{1,...,2m\}$). Such a state is characterized  through 
\begin{eqnarray*}
\omega(B^\ast(f)B(g))=(f,Sg)
\end{eqnarray*}
by its 2-point operator $S$ on ${\mathcal K}$ with the properties (see \cite{A71})
\begin{eqnarray*}
0\le S\le 1,\quad S+JSJ=1.
\end{eqnarray*}

\vspace{0.5cm}

{\it Remark 9}\,\,The definition of $B(f)$ in \eqref{def:B} and of $J$ after \eqref{prop:B}, with $f\in {\mathcal K}=\hh\oplus\hh$,  is a special case of a self-dual CAR algebra, and the non-equilibrium steady state $\omega_+$ from \eqref{ness} is a quasifree state in the sense \eqref{qf1}, \eqref{qf2}, see \cite{AP03}. 

\vspace{0.5cm}

Let $c_i$, $i=1,...,n$,  be any fermions in $\mA_n\simeq\mL(\CAR(\hh))\simeq \C^{2^n\times 2^n}$, 
\begin{eqnarray}
\label{c:car}
\{c_i,c_j\}=0,\quad \{c_i^\ast,c_j\}=\delta_{ij},
\end{eqnarray} 
and define operators $e_{\alpha\beta}^{(i)}$, $i=1,...,n$, $\alpha,\beta=1,2$, by
\begin{eqnarray}
\label{def:e}
e_{11}^{(i)}=c_i^\ast c_i,\quad e_{12}^{(i)}=c_i^\ast,\quad e_{21}^{(i)}=c_i,\quad e_{22}^{(i)}=c_i c_i^\ast. 
\end{eqnarray} 
It follows from the CAR \eqref{c:car} that the operators $e_{\alpha\beta}^{(i)}$ have  the properties
\begin{eqnarray}
\label{com1}
e_{\alpha\beta}^{(i)}\,e_{\gamma\delta}^{(j)}&=&(-1)^{(\alpha+\beta)(\gamma+\delta)}\,e_{\gamma\delta}^{(j)}\,e_{\alpha\beta}^{(i)},\quad \mbox{for}\quad i\neq j,\\
\label{com2}
e_{\alpha\beta}^{(i)}\,e_{\gamma\delta}^{(i)}&=&\delta_{\beta\gamma}\,e_{\alpha\delta}^{(i)}.
\end{eqnarray}

\vspace{0.5cm}

{\it Remark 10}\,\,Note that, due to \eqref{B:car}, the operators $c_i=B(f^{(i)})$ are fermions \eqref{c:car}, if and only if $(Jf^{(i)},f^{(j)})=0$ and $(f^{(i)},f^{(j)})=\delta_{ij}$. In the case at hand, the fermions are given in \eqref{def:bn} (there is a $^\ast$-isomorphism between the CAR algebra $\mA$ over $\hh$ and the self-dual CAR algebra over $\hh\oplus\hh$, see  \cite{A71}).

\vspace{0.5cm}

\begin{lemma}
\label{lemma:factorizing}
Let $e_{\alpha\beta}^{(i)}$ be defined by \eqref{def:e} for fermions $c_i=B(f^{(i)})$, and let $\omega$ be a quasi-free state on $\mA_n$ which satisfies
\begin{eqnarray}
\label{factorizing1}
\omega(c_ic_j)=0,\quad \omega(c_i^\ast c_j)=\delta_{ij}\,\omega(c_i^\ast c_i),\quad  i,j=1,...,n.
\end{eqnarray}
Then, $\omega$ factorizes on $\prod_{i=1}^ne^{(i)}_{\alpha_i\beta_i}$ as
\begin{eqnarray}
\label{factorizing2}
\omega\!\left(\prod_{i=1}^ne^{(i)}_{\alpha_i\beta_i}\right)=\prod_{i=1}^n\delta_{\alpha_i\beta_i}\,\omega(e^{(i)}_{\alpha_i\alpha_i}).
\end{eqnarray}
\end{lemma}

{\bf Proof}\,\, Assume first that the number of $B(\cdot)$ in $\prod_{i=1}^ne^{(i)}_{\alpha_i\beta_i}$ is odd. Then, there exists an  $i\in \{1,...,n\}$ such that $\alpha_i\neq\beta_i$ and, hence, due to \eqref{qf1}, \eqref{factorizing2} holds. Now, let the number of $B(\cdot)$ in $\prod_{i=1}^ne^{(i)}_{\alpha_i\beta_i}$ be even. If $\alpha_i\neq\beta_i$ for some $i\in \{1,...,n\}$, then every term in the sum \eqref{qf2} contains a factor of the form $\omega(B(J^{\epsilon_i}f^{(i)})B(J^{\epsilon_j}f^{(j)}))$ with $i\neq j$ and some $\epsilon_i,\epsilon_j\in\{0,1\}$. Therefore, due to \eqref{factorizing1}, we have
\begin{eqnarray}
\label{prefactorizing}
\omega\!\left(\prod_{i=1}^ne^{(i)}_{\alpha_i\beta_i}\right)=\prod_{i=1}^n\delta_{\alpha_i\beta_i}\,\,\omega\!\left(\prod_{j=1}^ne^{(j)}_{\alpha_j\alpha_j}\right).
\end{eqnarray}
For the same reason, applying \eqref{qf2} to the right hand side of \eqref{prefactorizing}, the only nonvanishing term is the one for which $\pi$ is the identity permutation. This implies \eqref{factorizing2}. \hfill $\Box$

\begin{figure}[h!]
\setlength{\unitlength}{1cm}
\begin{center}
\centering
\begin{picture}(7,2)
\multiput(0,0)(0.5,0){4}{\circle*{0.15}}
\put(0.25,0){\oval(0.5,1.5)[t]}
\put(1.25,0){\oval(0.5,1.5)[t]}
\put(1.85,0.25){$-$}
\multiput(2.5,0)(0.5,0){4}{\circle*{0.15}}
\put(3,0){\oval(1,1.5)[t]}
\put(3.5,0){\oval(1,1.5)[t]}
\put(4.25,0.25){$+$}
\multiput(5,0)(0.5,0){4}{\circle*{0.15}}
\put(5.75,0){\oval(1.5,1.5)[t]}
\put(5.75,0){\oval(0.5,1.5)[t]}
\end{picture}
\end{center}
\vspace{-0.5cm}
\caption{The pairings of the right hand side of \eqref{prefactorizing} for $n=2$.}
\end{figure}
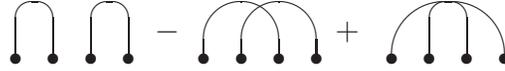

\vspace{0.5cm}

\begin{lemma}
\label{lemma:onb}
Let $e_{\alpha\beta}^{(i)}$ be defined by \eqref{def:e} for fermions $c_i=B(f^{(i)})$. Then, the set of operators $\prod_{i=1}^ne_{\alpha_i\beta_i}^{(i)}$, $\alpha_1,\beta_1,...,\alpha_n,\beta_n=1,2$,  is an orthonormal basis of the Hilbert space $\mA_n$ with scalar product $(A,B)\mapsto \tr(A^\ast B)$.
\end{lemma}

{\bf Proof}\,\, We show that the $2^{2n}$ vectors $\prod_{i=1}^n e_{\alpha_i\beta_i}^{(i)}$ with $\alpha_1,\beta_1,...,\alpha_n,\beta_n=1,2$, are orthonormal with respect to the scalar product  $\mA_n\times\mA_n\ni (A,B)\mapsto \tr(A^\ast B)$  by proving
\begin{eqnarray}
\label{ons}
\tr\,\left(\prod_{i=1}^n e_{\alpha_i\beta_i}^{(i)}\left(\prod_{j=1}^n e_{\gamma_j\delta_j}^{(j)}\right)^\ast\right)=\prod_{i=1}^n\delta_{\alpha_i\gamma_i}\delta_{\beta_i\delta_i}.
\end{eqnarray}
For this purpose, we push the $j$-th term in the second product on the left hand side of \eqref{ons} to the left using \eqref{com1} until it it hits the $j$-th term in the first product and apply \eqref{com2} (note that $e_{\gamma_j\delta_j}^{(j)\ast}=e_{\delta_j\gamma_j}^{(j)}$). Doing so, from $j=n$ until $j=1$, we arrive at
\begin{eqnarray}
\label{prod}
\prod_{i=1}^n e_{\alpha_i\beta_i}^{(i)}\left(\prod_{j=1}^n e_{\gamma_j\delta_j}^{(j)}\right)^\ast=\prod_{i=1}^n\delta_{\beta_i\gamma_i}\prod_{j=1}^{n-1}(-1)^{(\gamma_j+\delta_j)\sum_{k_j=1}^n(\alpha_{k_j}+\delta_{k_j})}\,\prod_{l=1}^n e_{\alpha_l\delta_l}^{(l)}.
\end{eqnarray}
Since $\tr(\cdot)/2^n$ is a quasi-free state on $\mA_n$ (see for example \cite{A71}) which, due to the CAR \eqref{c:car} and the cyclicity of the trace, satisfyies \eqref{factorizing1}, we have, using \eqref{factorizing2} and \eqref{prod}, 
\begin{eqnarray*}
\tr\,\left(\prod_{i=1}^n e_{\alpha_i\beta_i}^{(i)}\left(\prod_{j=1}^n e_{\gamma_j\delta_j}^{(j)}\right)^\ast\right)=\prod_{i=1}^n\delta_{\beta_i\gamma_i}\prod_{j=1}^{n-1}(-1)^{(\gamma_j+\delta_j)\sum_{k_j=1}^n(\alpha_{k_j}+\delta_{k_j})}\,\prod_{l=1}^n \delta_{\alpha_l\delta_l}\,\tr(e^{(l)}_{\alpha_l\alpha_l}) 
\end{eqnarray*}
(of course, \eqref{factorizing2} can easily be verified directly for $\tr(\cdot)/2^n$). Finally, using again the CAR \eqref{c:car} and the cyclicity of the trace, we have $\tr(e^{(l)}_{\alpha_l\alpha_l})/2^n=1/2$ and \eqref{ons} follows. \hfill $\Box$

\section{Toeplitz operators}
\label{app:Toeplitz}

{\bf Toeplitz operators} \cite[p.185]{BS99}\,\,  Let $N\in\N$. We
define the space $l^2_N$ of all $\C^N$-valued sequences $f=\{f_i\}_{i=1}^\infty$, $f_i\in\C^N$, by
\begin{eqnarray*}
l^2_N=\{f:\N\to\C^N\,|\, \|f\|<\infty\},\quad
\|f\|=\left(\sum_{i=1}^\infty\|f_i\|^2_{\C^N}\right)^{1/2},
\end{eqnarray*}
where $\|\cdot \|_{\C^N}$ denotes the Euclidean norm on $\C^N$. For $N=1$ we
write $l^2\equiv l^2_1$. 

Let $\{a_x\}_{x\in\Z}$ be a sequence of $N\!\times\! N$ matrices, $a_x\in\C^{N\times N}$. The Toeplitz operator defined through its action on elements of $l^2_N$ by
$f\mapsto \{\sum_{j=1}^\infty a_{i-j} \,f_j\}_{i=1}^\infty$ is a bounded operator on $l^2_N$, if and only if
\begin{eqnarray*}
a_x=\int_0^{2\pi}\!\frac{d\xi}{2\pi}\,a(\xi)\,e^{-ix\xi}
\end{eqnarray*}
for some $a\in L^\infty_{N\times N}$ \,(see \cite[p.186]{BS99}), where we define (with $\T=\{z\in\C\,|\,|z|=1\}$)
\begin{eqnarray*}
L^\infty_{N\times N}=\{\phi:\T\to\C^{N\times N}\,|\,
\phi_{ij}\in L^\infty(\T), \,i,j=1,...,N\}.
\end{eqnarray*}
In this case, we write the Toeplitz operator as
\begin{eqnarray*}
T[a]=\left[
    \begin{array}{lllll}
    a_0 & a_{-1} & a_{-2} & ...\\
    a_1 & a_0 & a_{-1} & ...\\
    a_2 & a_1 & a_0 &... \\
    ... & ... & ... & ...\\
    \end{array}\right].
\end{eqnarray*}
The function $a\in L^\infty_{N\times N}$ is called the symbol of $T[a]$. If $N=1$ the symbol $a\in L^\infty\equiv L^\infty_{1\times 1}$ and the Toeplitz operator $T[a]$ are called scalar, whereas for $N>1$ they are called block.

For $n\in\N$, let $P_n$ be the projections on $l^2_N$,
\begin{eqnarray*}
 P_n(\{f_1,...,f_n,f_{n+1},...\})=\{f_1,...,f_n,0,0,...\}.
\end{eqnarray*}
With the help of these $P_n$, we define the truncated $Nn\!\times\!Nn$ Toeplitz matrices as 
\begin{equation*}
T_n[a]= \left.P_n T[a] P_n\right|_{\mathrm{ran\,}P_n},
\end{equation*}
where ran$A$ denotes the range of the operator $A$.

\vspace{0.5cm}

{\bf Norm of Toeplitz operators} \cite[p.186]{BS99}\,\, The norm of a Toeplitz operator is related to its symbol,
\begin{eqnarray}
\label{ToeplitzNorm}
\|T[a]\|=\|a\|_\infty,
\end{eqnarray}
where $\|a\|_\infty$ is defined to be the operator norm of the multiplication operator acting on $\oplus_{1}^NL^2(\T)$ by multiplication with the matrix function $a\in L^\infty_{N\times N}$. We have 
\begin{eqnarray}
\label{def:phi_infinity}
\|a\|_\infty=\mbox{ess} \sup_{\hspace{-0.5cm}\xi\in
[0,2\pi]}\,\|a(\xi)\|_{\mL(\C^N)},
\end{eqnarray}
where $\|\cdot\|_{\mL(\C^N)}$ is the operator norm induced by the $l^2$ norm on $\C^N$, see also \cite[p.101]{BS90}.

\vspace{0.5cm}

{\bf Szeg\"o's first limit theorem in the block case} \cite[p.202]{BS99}\,\, Let $a\in L_{N\!\times\! N}^\infty$ be self-adjoint, $a^\ast=a$, and let $\lambda_1^{(n)},...,\lambda_{Nn}^{(n)}$ be the eigenvalues of the block Toeplitz matrix $T_n[a]$. Then, for any continuous function $f$ with compact support, $f\in C_0(\R)$, we have
\begin{eqnarray}
\label{szego}
\lim_{n\to\infty}\frac{1}{Nn}\sum_{k=1}^{Nn}f(\lambda_k^{(n)})=\frac{1}{N}\int_0^{2\pi}\!\frac{d\xi}{2\pi}\,\tr f(a(\xi)).
\end{eqnarray}

\vspace{1cm}

\textbf{Acknowledgements}\quad It is a great pleasure to thank Herbert Spohn for his interesting comments. Moreover, I am grateful to the referee for his constructive remarks. 



\end{document}